\documentclass[12pt]{article}
\usepackage{algorithm}
\usepackage{algpseudocode}
\usepackage{lipsum}
\usepackage{amsmath}
\usepackage{amsfonts}
\usepackage{amssymb}
\usepackage{graphicx}
\usepackage{subfigure}
\usepackage{cite}
\usepackage{amsthm}
\usepackage{epstopdf,epsfig}
\usepackage{xcolor}
\usepackage{balance}
\usepackage{colortbl}
\usepackage{url}
\usepackage{algpseudocode}
\usepackage{booktabs}
\usepackage{amsmath,amsfonts,amsthm}

\usepackage{soul}
\usepackage{csquotes}

\newtheorem*{theorem*}{Theorem}
\newtheorem{mydef}{Definition}
\usepackage{algorithm}
\usepackage{tabularx}
\usepackage{natbib}

\begin{document}

\title{Fostering Peer Learning through a New Game-Theoretical Approach in a Blended Learning Environment}

\author{Seyede~Fatemeh~Noorani$^a$,Mohammad~Hossein~Manshaei$^a$,\\~Mohammad~Ali~Montazeri$^a$, Behnaz Omoomi$^b$\\$^a$Department of Electrical and Computer Engineering
\\$^b$Department of Mathematical Sciences
\\Isfahan University of Technology
\\Isfahan 84156-83111, Iran
\\E-mails: sf.noorani@ec.iut.ac.ir, \\ \{manshaei, montazeri,bomoomi\}@cc.iut.ac.ir}

\maketitle
\begin{abstract}
Obtaining knowledge and skill achievement through 
peer learning- can lead to higher academic achievement.
However, peer learning implementation is not just about putting students together and hoping for the best. At its worst-designed, peer learning may result in one person doing all the effort for instance, or may fail to encourage  the students to interact enough with the task and so enhance the task in hand. 

This study proposes a mechanism as well as an instructional design to foster well-organized peer learning based on game theory $(PD\_PL)$. The proposed mechanism uses prisoner's dilemma
and maps the strategy and payoff concepts found in prisoner's dilemma onto a peer learning atmosphere.  

$(PD\_PL)$ was implemented during several sessions of four university courses and with 142 computer engineering students.  
The study results indicated that $PD\_PL$ was beneficial and favourable to the students. 
Further analysis showed that the $PD\_PL$ had sometimes even enhanced learning by up to $47.2\%$. 


\end{abstract}

\section{Introduction}
\label{sec:into}
Peer learning(PL) can be defined as obtaining knowledge and skill achievement through learning among status equals \cite{RN27}. In this method, people of the same social group work as amateur instructors in pairs and help each other to teach and learn from each other \cite{RN28}. The students who learn via this method have a better understanding of the content of the lesson, have higher motivation, and learn faster \cite{RN35}. Given the positive research results of peer learning, it would make sense to harness these benefits by designing the classroom activities that lead students to peer learning \cite{RN31, whitman1988peer}. 

However, peer learning is not putting students together and hoping for the best. For instance, it may result in one person doing all the work or may fail to lead the students to engage enough interaction and enhance the task in hand consequently; hence the need for a well-designed structure \cite{RN27}. This paper proposes a novel game theoretical mechanism for implementing peer learning.

With the aid of mathematical models, game theory analyzes the method of cooperation or conflict between intelligent and rational decision makers \cite{myerson1991game}.  In this atmosphere, each decision maker tries to increase their payoff through interaction with other decision makers. Today many applications of game theory habe been reported in various fields. For example, we can refer to cases such as auctions, bargaining and collective decision-making in economies, and voting in the fields of politics and security as well as privacy in the context of networking. Besides, as Cohen  et al. \cite{cohen2018student} state, game theory can be applied in the area of education to enhance learning results. For instance, it has been applied to investigate the effect of strict or lenient scoring on increased or decreased effort of students or instructors \cite{Correa87}, to study the impact of the number of students in a class on their success \cite{RN2,RN3}, to model methods of effort-making by instructors and students \cite{RN4}, to model the interactions of instructor and student \cite{RN6}, to model the participatory or competitor behavior between instructors \cite{RN5}, to evaluate a student's cooperative behavior \cite{RN7,RN12,su2018individuals}, to develop digital game-based learning for the concept of prisoner's dilemma\cite{moniaga2017prototype}, and to create a competition among students \cite{RN16}. Our study is different from the mentioned works because in fact, our contribution is applying prisoner's dilemma (PD)-which is a famous instance of game theory- to prepare a peer learning environment and motivate the students to actively participate in their stages of learning.

In peer learning, a successful effort by both of the students results in increased learning outcomes, but an unsuccessful effort by both parties results in lower learning outcomes. In this method, the non-contributing students are referred to as free riders, who enjoy the benefits of group activities without participation and responsibility in the group work. This behavior of students in peer learning is the same as that of the player in prisoner's dilemma. Prisoner's dilemma shows how it's possible for two rational people to tend not to cooperate (defect) despite their cooperation possibly leading to a higher payoff for both of them. In PD, the payoff of a participant who puts in more effort will be less than the one who makes less effort. In this case, the person who makes less effort is an example of a free rider. On the other hand, in PD, effort is made to lead participants towards having better and greater cooperation. As in PD, greater effort made by both participants in peer learning results in increased learning improvement, which is the main goal of our suggested mechanism. 

In this paper, we first present a novel mechanism based on peer learning and prisoner's dilemma. Secondly, as in a PD situation where the cooperation of both participants leads to a better result, our proposed mechanism tries to encourage the students to make a greater effort to enhance learning their own achievement. Finally, we demonstrate that the presented mechanism could enhance learning outcomes.

To investigate the effect of our mechanism on learning performance, it has been implemented on four groups of students in different courses. A statistical test has been used to analyze the results of pre-test and post-test exams taken before and after mechanism implementation respectively. Since the test contains requirements such as data imputation, we will also explain how to utilize and implement the mechanism and how to analyze the results, in the section devoted to our methodology. The result of its implementation indicates that the proposed mechanism has a positive impact on personal learning outcomes.

The rest of the paper is organized as follows. After introducing peer learning and game theory in section \ref{sec:LT}, a literature review of the application of game theory concerning learning will be shown in Section \ref{sec:RelatedWork}. In Section \ref{sec:TheproposedMechanism}, the suggested mechanism will be presented. Section \ref{sec:methodology} covers its method of implementation and data collection as well as the method of analyzing the received data. The result of the analysis, as well as the procurement of the mechanism, will be discussed in Section \ref{sec:results}. In the final sections, a discussion and conclusion are drawn, and future works are presented.

\subsection{Research Questions}
This study particularly attempted to answer the following research questions:

\begin{itemize}
\item Learning improvement: Does the proposed peer learning mechanism, enriched by game theory, enhance students' learning outcomes? 
\item Free rider prevention: Is PD\_PL able to stop the free rider problem?
\item Subjective evaluation: How do students evaluate the PD\_PL process?
\end{itemize}

\section{Background}
\label{sec:LT}
Since the design of the suggested mechanism is based on game theory and peer learning, we briefly review these concepts in this section.

\subsection{Peer Learning}
\label{PL}
Peer learning is an educational practice in which students are able to reach their goals by working together \cite{RN28}. This method of learning causes the students to not only rely on the instructor and the syllabus of the book, but also to discuss every opinion concerning themselves as well as others in the same group. On the other hand, students can easily discuss their opinions with others in the absence of the instructor. Another point is that questions and answers can be analyzed from several points of view other than those of the instructors \cite{RN31}.

Topping et al. \cite{RN30} classify the advantages of peer learning from the viewpoints of students, instructors, and the educational system. We will delve into some of them in more details as follows:

Advantages of the peer learning from the students' point of view:
\begin{itemize}
\item Higher academic achievement
\item Improvement in interpersonal relationships
\item Improvement in individuality and society (For instance, more feeling of self-worth, more positive attitudes to institute and learning)
\item More optimistic learning atmosphere
\item Motivational improvement (for example, more pleasurable and better opportunities to socialize with peers)
\end{itemize}

Advantages of peer learning from the instructors' point of view:
\begin{itemize}
\item Instructional development (For instance, raising educational services and observing individual student performance)
\item Classroom management (For example, reducing unsuitable academic and social behavior, opportunities to teach new appropriate classroom behaviors)
\item Simple implementation
\end{itemize}

Advantages of peer learning from the educational system point of view:
\begin{itemize}
\item A host of strategies for improving student achievement
\item A means of raising educational reforms
\item A collection of interpositions to simplify addition, improve general classroom discipline, and avoid academic failure
\end{itemize}

As the proposed mechanism aims to increase learning through teammates, peer learning considerations should be taken into account. Topping \cite{RN29} lists the requirements to be clarified in peer learning. He states that the context, goals, curriculum area, participants, helping technique, contact method, materials, participants and their training technique, mechanism of process monitoring, assessment of students' tactics, evaluation method, and how to get participants' feedback should be determined. In the section devoted to our methodology (i.e., Section~\ref{sec:methodology}), we will explain how to implement such specifications in our mechanism.

\subsection{Game Theory and Prisoner's Dilemma}
Game theory forms a mathematical model of cooperation and conflict between intelligent and rational decision makers \cite{myerson1991game}. A rational decision maker tries to maximize his payoff against other decision makers. The decision maker in game theory is known as the player. In each game, players work together and in every stage of the game, choose a strategy from their strategy set. The players take their payoff with respect to the payoff matrix.

Nash Equilibrium and Pareto Efficiency (Pareto Optimum) are two fundamental definitions in game theory, that we explain as follows.
\begin{mydef}
	Strategy profile $s^*$ constitutes a Nash Equilibrium (NE) if each player $i$ meets the below condition:
	\begin{equation}
	\label{eq:nash}
		u_i(s^*_i, s^*_{-i})  \geq u_i(s_i, s^*_{-i}), \forall  s_i \in S_i,
	\end{equation}
\end{mydef}

where, $s_i$ and $s_{-i}$ are strategy of player $i$ and strategies of remaining players respectively, and $u_i(s_i, s^*_{-i})$ is the payoff function of player $i$.

John Nash has shown that at least one mixed strategy Nash Equilibrium will exist for any finite game \cite{RN36}. Some points have been mentioned regarding NE:
\begin{itemize}
\item Since each player selects his/her best response to the other players' choices, NE can be seen as an outcome of mutual best responses.
\item The NE definition states that, no player can increase his/her payoff by deviating unilaterally.
\item Accordingly, no player regrets his/her action when they play in a NE.
\end{itemize}
\begin{mydef}
A strategy profile of a game is (weakly) pareto efficient iff there is no other strategy profile that would make all players better off \cite{myerson1991game}.
\end{mydef}

In the proposed mechanism, we apply prisoner's dilemma, which is a famous game in game theory. It models the situation where two people who display rational behavior and know that cooperating together for the same subject would be beneficial to both of them, but prefer not to cooperate. In this game, there are two players (Player 1 and Player 2) and each one has a strategy set \{Cooperate, Defect\}. Table ~\ref{tab:PD} shows the payoff matrix of this game in which\cite{RN26}:
\begin{center}
$a<b<c<d$ 
\end{center}
and
\begin{center}
$\dfrac{(a+b)}{2}<c$
\end{center}
\begin{table}[htp]
\caption{Payoff Matrix of Prisoner's Dilemma}
\begin{center}
\begin{tabular}{c}
\includegraphics[scale=0.9]{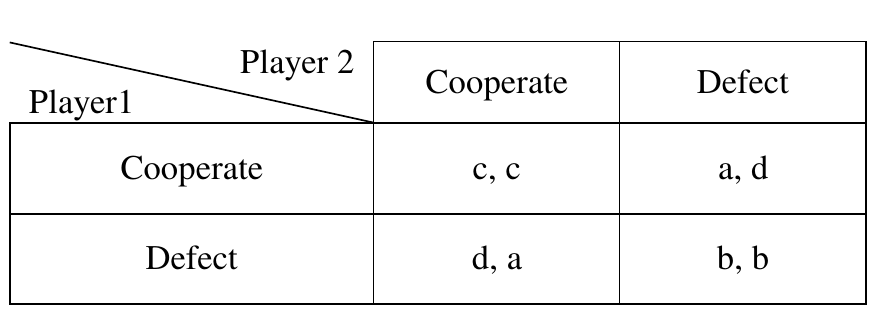}
\end{tabular}
\end{center}
\label{tab:PD}
\end{table}

The rows of this table show the strategies of the first player and the columns show the strategies of player 2. There are two values in each cell of the payoff matrix. For instance, if the strategy profile (Cooperate, Defect) is chosen - that is, player 1 chooses  \enquote{Cooperate} strategy and the player 2 chooses  \enquote{Defect} - then the payoff of the player 1 is  \enquote{a} and that of the player 2 is  \enquote{d}.

In this game, if both players choose the strategy \enquote{Cooperate}, the payoff of both players are \enquote{c}. This payoff is higher than what players gained by choosing strategy \enquote{Defect}. While one player chooses \enquote{Cooperate} and the other chooses \enquote{Defect}, the payoff of the player who chooses \enquote{Cooperate} is less than the one who chooses \enquote{Defect}. For example, if player 1 chooses to cooperate and player 2 chooses not to cooperate, player 1 obtains payoff \enquote{a} and other gains payoff \enquote{d}, and as we mentioned, \enquote{a} is lower than \enquote{d}.

In the prisoner's dilemma, NE is the strategy profile (Defect, Defect). This is due to the fact that none of the players gain more payoff by changing strategy unilaterally. For example, if the first player chooses the \enquote{Cooperate} strategy instead of \enquote{Defect}, he receives \enquote{a} which is lower than \enquote{d}. The same condition will happen to the other player.

In this game, the strategy profile (Cooperate, Cooperate) is a pareto efficiency. That is, the other strategy profile does not contribute any extra payoff to both players. As an example, by choosing a different strategy profile (Cooperate, Defect), player 2 will receive a higher payoffو but the payoff of the first player will be lower. 

We will explain the relationship between the suggested mechanism and prisoner's dilemma more in Section \ref{sec:TheproposedMechanism}.

\section{Related Work}
\label{sec:RelatedWork}
By reviewing the related research on using game theory in a learning context, we divide these studies into five groups. Some articles model the interaction between instructor and student \cite{Correa87,RN2,RN3,RN4,RN6,RN10,RN11,colman2018persistent}. Some other works model the interaction between instructors \cite{RN5}. Beside these articles, there are also articles that use game theory in evaluating and commentating on students' cooperation behavior \cite{RN7,RN12,RN13,RN14,RN9}. Another group of articles use game theory to train cooperative behavior \cite{RN15}. Finally, \cite{RN16,noorani2018game} applies game theory to increase competition between students. 

In the following, we explain each group in detail.
\begin{itemize}
\item
\textbf{Using game theory in modeling instructors' and students' interactions}: Between 1987 to 2003, Hector Correa procured a series of works for using game theory in the area of education. Initially, he was theoretically examining the use of economic theory in education \cite{Correa87}. Then in \cite{RN2, RN3}, he investigated the correlation between the number of students in a class and educational achievement. He indicated that as the number of students in a class rises, the rate of success of the students would be reduced. These articles had not been implemented. 

In 2003, Correa also modeled the interaction between instructors and students \cite{RN6}. In this study, instructors and students were divided into capable/incapable and hardworking/lazy groups. The achievement function of students was formed as $h=a_{1}t_{s}+a_2t_st_c+a_3t_st_g-a_4{t_c}^2-a_5{t_c}^2-a_6{t_g}^2$, in which $t_s$ and $t_c$ denoted the time allocated by a student and an instructor for a lesson respectively. The value of $t_g$ was the time allocated by an instructor to each student individually (as an example, the required time for marking the exam paper of a student). Finally, he applied simulation to calculate the values of $a_1$ to $a_6$ in the NE situation. 

Moga et al. \cite{RN10} defined the strategy set \{Cooperative, Non-Cooperative\}  for students, and \{Using the classic teaching method, Using interaction model in teaching\} for instructors. Then, they determined the payoff matrix and NE in order to choose the best educational policy.

Oltean et al. \cite{RN11} involved the financial outcome. They defined the strategy set \{Study/ Don't study\} for students, and \{Verify/ Don't Verify\} for instructor. In the \enquote{Verify}, the instructor assessed the students via very precise testing. In the \enquote{Don't Verify}, the financial outcome of the institute was important, and the instructor had a more lenient approach to evaluating students. Instructors chose their strategy with respect to the value of three parameters (Professional Effort/Risk of Losing Money/Professional Prestige). On the other hand, students choose their strategies in respect to the parameters (Knowledge/Risk of losing the Degree/Professional Effort). Then the quality values (H, h, l, L) for the mentioned parameters were considered, in which $H>h>l>L$. Finally, after defining the NE of each scenario, it was calculated.

\item
\textbf{Applying game theory in modeling the interaction between instructors:} Correa analyzed cooperative and competitive behavior between instructors at an educational institute \cite{RN5}. He defined $h_i=n_i \times s_i$ as an achievement function of the instructor, in which $n_i$ was the number of students who enrolled in the lesson of instructor $i$, and $s_i$ was the number of students who successfully passed the lesson.
 
Two instructors could cooperate so that they sought to maximize $h=h_1+h_2=n_1s_1+n_2s_2$. In another approach, they could adopt competitive behavior so that finding the NE could solve the problem. 

\item
\textbf{Applying game theory in evaluating and analyzing cooperative behavior:} Waddell et al. \cite{RN7} used PD to analyze the cooperative or competitive behavior of students when they play PD against a known or unknown person. Hemesath \cite{RN12}  investigated the effect of nationality, gender and familiarity in PD. Molina and et al.  in \cite{RN14} studied the relation between the gender of students and their strategy selection in PD. Gray et al. examined the effects of age on cooperation\cite{gray2017game}. Fernndez-Berrocal \cite{RN13} evaluated the relation between emotional intelligence and making decisions in choosing a cooperative or competitive strategy in iterated PD. It concluded that people with higher emotional intelligence make more effective decisions and they thought about long term gains. 

Chiong et al. \cite{RN9} used evolutionary game theory to analyze the cooperation behavior of players in group work. In their experiment, active players of non-active groups were transferred to active groups and non-active players were transferred to non-active groups. The result of their experiment showed that the active players remained active and non-active players remained non-active. 

\item
\textbf{Applying game theory to train cooperative behavior:} Fan \cite{RN15} used PD to train cooperative behavior. In this study, participants were grouped in pairs. Each participant was given two cards; one marked with a triangle and one with a circle. Each team member showed a card to their teammate simultaneously. Table ~\ref{tab:ShowingCards} indicates the payoff matrix of this game. After some sessions, the experimenter gives a short lecture. This lecture (shown in Figure ~\ref{fig:ShowingCards2})  was designed as a treatment variable by which students were explicitly told that it was a good thing to cooperate. The research showed that the proportion of cooperative individuals increased meaningfully immediately after the lecture.
\begin{table*}[htp]
\caption{Payoff Matrix of showing cards game\cite{RN15}}
\begin{center}
\begin{tabular}{c}
\includegraphics[scale=0.8]{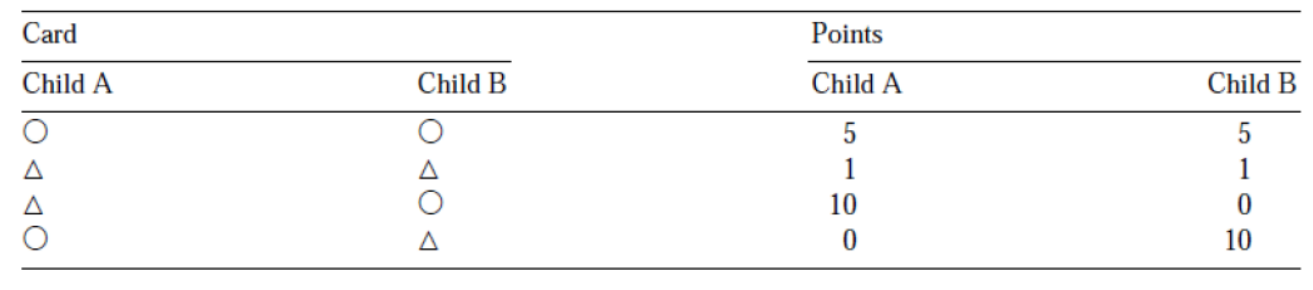}
\end{tabular}
\end{center}
\label{tab:ShowingCards}
\end{table*}

\begin{figure}[htp]
	\centering
	\includegraphics [scale=0.8]{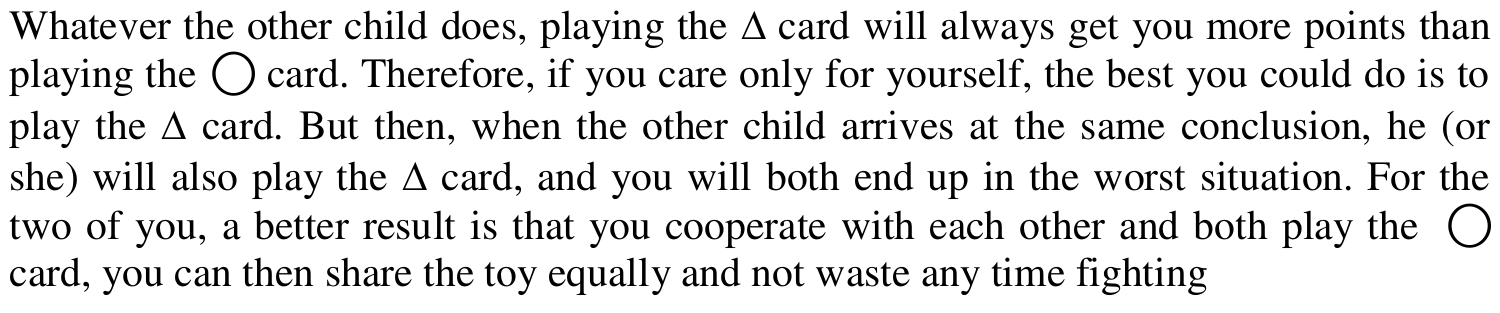}
	\caption{The essence of the lecture given by experimenter \cite{RN15}}
	\label{fig:ShowingCards2}
\end{figure}

\item
\textbf{Applying game theory to motivate competition:} Burguillo \cite{RN16} used PD to motivate competition between students in order to enhance their java programming skills. The students should use java to create a network program to run PD. Article \cite{noorani2018game} proposed a game theoretical approach to stimulate learners to take part in a competition to provide more useful explanations.
\end{itemize}
\subsection{How does PD\_PL differentiate from related work?}
There are distinctions between $PD\_PL$ and the mentioned studies. First, $PD\_PL$'s main goal is to enhance learning using peer learning. Second, it models the interaction between students using prisoner's dilemma. Thirdly, $PD\_PL$ uses PD's payoff matrix to determine students' scoring. Finally, it was implemented and the result was reported; that is, this study has not relied solely on simulation or theoretical proof. Future, $PD\_PL$ applies PD to peer learning, while article \cite{RN16} defines PD as a java programming exercise, and does not use a mathematical model of game theory.

\section{The Proposed Mechanism}
\label{sec:TheproposedMechanism}
We aim to design a mechanism based on game theory to enhance learning outcomes. As mentioned above, we apply prisoner's dilemma to the proposed mechanism. In PD, strategy profiles (Defect, Defect) and (Cooperate, Cooperate) are NE and pareto optimum respectively. The favourite situation is motivating players to move from NE to pareto optimum to gain higher payoffs. The same holds true in peer learning in which the effort of both students results in greater learning outcomes. We use the payoff as a reward to motivate students to make all-out efforts. Thus, we tell the students that the total payoff they gained will be taken into account for their mid-term score. Therefore, the mechanism motivates the students to make more effort to gain higher payoffs. 

On the other hand, we run the mechanism during several sessions. Therefore, if a student even faces a free rider, he can negotiate with the teammate to make more effort or can even change the teammate in the next session. Following, we describe the proposed mechanism in detail.

\subsection{PD\_PL: A mechanism based on prisoner's dilemma and peer learning}

\begin{figure*}[tp]
	\includegraphics [scale=1]{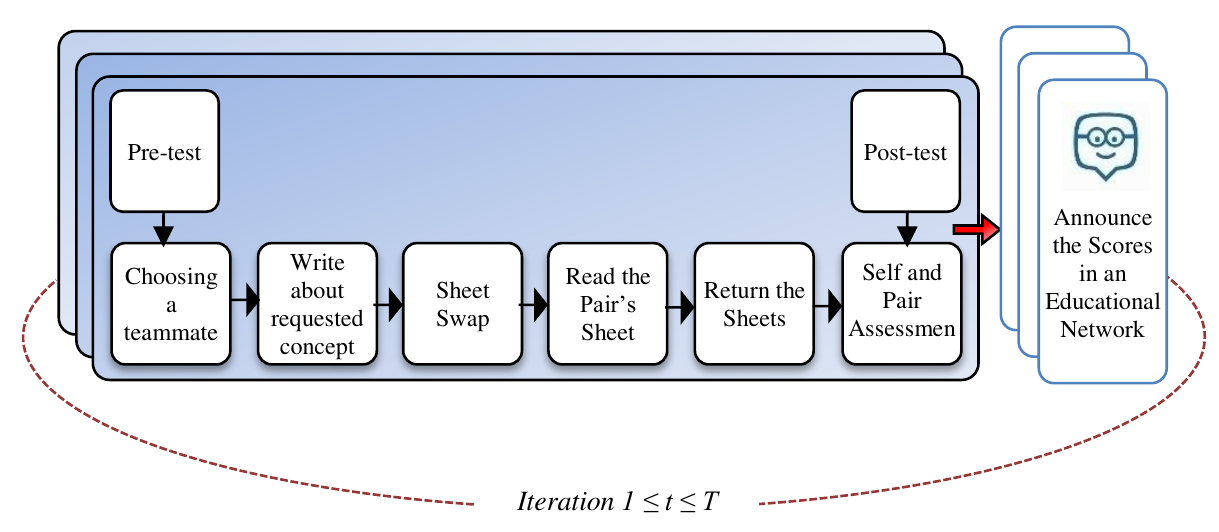}
	\centering
	\caption{The stages of PD\_PL}
	\label{fig:ProposedMechanism}
\end{figure*}

Figure ~\ref{fig:ProposedMechanism} illustrates the stages of PD\_PL. In each session where the PD\_PL is run, we ask the students to form a group of two students at their sole discretion. The mechanism is run at the end of some sessions after the instructor teaches the lesson. We give a sheet to students and ask them to briefly write about a given concept.  This concept arises out of the lesson taught in that session; the writing of which does not require more than 5 to 10 minutes.

We emphasize to the students that the written text must be understood by their teammate and that they should help their teammate to eliminate any probable misunderstanding.
After the time determined by the instructor elapses, we ask the students to swap their sheet with their partner. Then, we ask students to study theie teammate's sheet. Afterward, the students should return the sheet to the owner. We ask students to fill in the requested information at the bottom of the sheet in order to make sure they have studied their partner's sheet. The requested items include the assessment of their teammate's sheet, a self-assessment and an assessment of familiarity with the teammate. 

To evaluate the effect of PD\_PL on learning improvement, we use a pre-test before and a post-test after the mechanism's implementation. Pre-test and post-test are at the same degree of difficulty and both are related to the concept that we ask students to write about on the sheet.

We run the mechanism during several sessions. An important point is that the students do not know in which sessions the mechanism is run.

At the end of each PD\_PL execution, the score of each student is calculated. This score is based on the student's and the teammate's sheets. We will further explain about score calculation. 

After each execution of mechanism, we place the scores in Edmodo, that is an educational network. People could register as an instructor, student, or parent and use different facility of this educational network. Some facilities are creating a new course, leaving and responding messages publicly and privately, making quizzes, preparing exams and exercises, observing the students' activities, and uploading the exercises. 

Knowing their own and their teammate's score, students may decide to change the teammate in the next session of mechanism execution, or make a decision about the amount of information that they write in their sheet.

\subsection{Scoring Method}
\label{sec:ScoringMethod}
As mentioned above, we use pre-test and post-test to evaluate the student knowledge before and after executing the mechanism. Maximum score of pre-test, post-test and the sheet in each session of PD\_PL implementation is 2. 

In PD\_PL, Equation (\ref{obtain_score}) is used to calculate the score of each member of a group in which $G_i$ and $G_j$ refere to score of sheet of student and his teammate, respectivly.

\begin{align}
\label{obtain_score}
\dfrac{G_i + G_j}{2}\times{1.2}
\end{align}

Table \ref{tab:ObtainedScores} shows some calculated scores concerning teammate sheets. 

\begin{table}[htp]
\caption{The scores gained by teammates based on their sheets' scores}
\begin{center}
\begin{tabular}{c}
\includegraphics[scale=1]{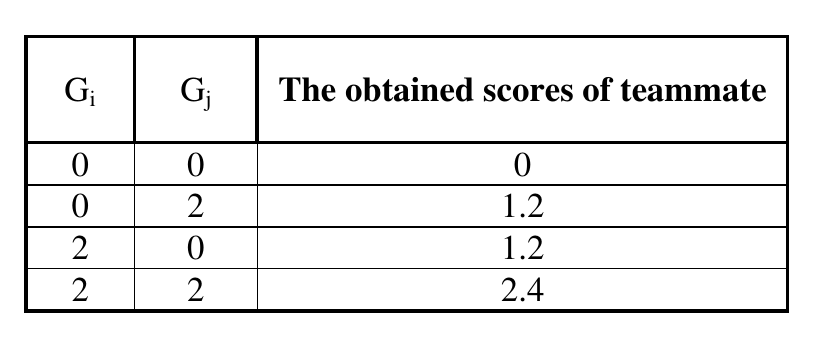}
\end{tabular}
\end{center}
\label{tab:ObtainedScores}
\end{table}

\subsection{Relation between Prisoner's Dilemma and PD\_PL}
Payoff is equal to benefit minus cost. As stated, if $G_i$ and $G_j$ are assumed to be the score of sheet of students $i$ and $j$ respectively, the score of each student is calculated according to Equation (\ref{obtain_score}). So the payoff for each student $i$ is benefit (that is calculated by Equation (\ref{obtain_score})) minus the cost ($G_i$). Hence, the payoff for student $i$ is $\dfrac{(G_i+G_j)}{2} \times 1.2 - G_i$, which is equal to $0.6 G_j- 0.4 G_i$. 
In the same way, the payoff of student $j$ is equal to $0.6 G_i- 0.4 G_j$ 

When one student writes more and the other (teammate) writes less, for example, $G_i > G_j$, we have a PD condition:

$G_i > G_j$  then $0.6G_i>0.6G_j$  and  $-0.4G_i < -0.4G_j $

We conclude $0.6G_i -0.4G_j > 0.6G_j -0.4G_i$, which means the payoff for student $j$ is higher than that for student $i$. It means the student (player) who makes the most effort gains a lower payoff than the student (player) who makes the least effort.

\begin{table}[htp]
	\caption{The payoff matrix of students in PD\_PL}
	\begin{center}
	\begin{tabular}{c}
	\includegraphics[scale=1]{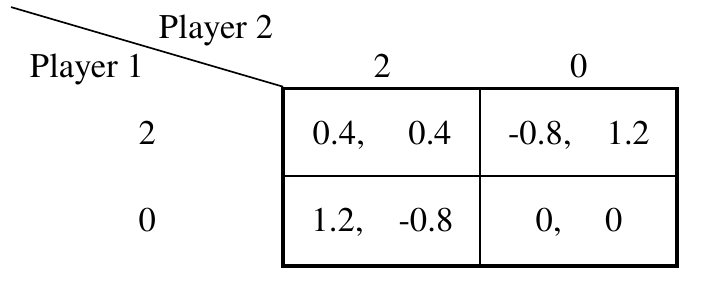}
	\end{tabular}
	\end{center}
	\label{tab:payOffExamples}
\end{table}

We also illustrate this condition with a simple numerical example in Table \ref{tab:payOffExamples}.
As an example, if the score of a student is zero and the value of his teammate is two, the score of each one is $(0+2)/2 \times 1.2=1.2$. Therefore, the payoff for the student is equal to $1.2-0=1.2$ and the payoff for their teammate is $1.2-2=-0.8$.

Comparing Table \ref{tab:payOffExamples} with Table \ref{tab:PD}, one can see that PD\_PL is in fact a prisoner's dilemma game. It means that when both students participate and cooperate and the score is $2$, their payoff is $0.4$, which is more than the situation where none of them makes any effort. In the case that one makes more effort and his sheet score is $2$ but his teammate obtains zero, the student's payoff is $-0.8$ and his teammate's payoff is $1.2$. That is, the student who makes more effort has a lower payoff than that of their teammate. 

As mentioned, we run the mechanism in several sessions. Therefore, even if a high-effort student faces a free rider, he can persuade their teammate to make more effort in the next few sessions. He can even change the teammate to achieve a higher payoff. Hence, PD\_PL motivates the students to make more efforts and achieve higher payoffs, resulting in higher learning achievement.

\subsection{Peer Learning Requirements in PD\_PL} 
In Section \ref{PL}, several requirements of peer learning were enumerated. Table \ref{tab:Implementation} shows how we meet these requirements in our mechanism. 

\begin{table}[htp]
\caption{Implementing peer learning requirements in PD\_PL.}
\begin{center}
\begin{tabular}{|m{2.5 cm}|p{12 cm} |} 
\hline
Context   & Mechanism is designed for the educational and learning area\\
\hline
Objectives & The goal is to improve learning \\
\hline
Curriculum area	& In addition to instructor training, mid-term exam and final exam, the PD\_PL facilitates learning improvement based on peer learning.\\
\hline
Participants	& The participants are a number of Computer Engineering students in courses  \enquote{Fundamentals of programming}, \enquote{Discrete mathematics}, and \enquote{Compiler principles techniques and tools}.\\
\hline
Helping technique	& Students help each other by writing their knowledge on sheets and exchanging their sheets. \\
\hline
Contact Method	& In every mechanism execution session, the students can contact each other. They can contact each other during the semester as well. This raises an opportunity for students to choose new teammate or persuade previous teammate to make more effort in order to obtain a higher payoff in the next iterations of the mechanism.\\
\hline
Materials & 	After each run of the mechanism, students need to use internet to observe their payoffs.\\
\hline
Training &	Before the start of the mechanism, students are trained about the method of mechanism and scoring method.\\
\hline
Process monitoring	& The quality of the proposed mechanism is controlled by comparing the scores of pre-tests and post-tests. The instructor monitors the mechanism execution as well. \\
\hline
Assessment of students tactic & Students assess teammates' and their own sheets every time the mechanism is executed.\\
\hline
Evolution method &	The effect of the mechanism is controlled by comparing post-test to pre-test results.\\
\hline
Feedback	&Every time the mechanism is run, students can provide their opinions about the method and quality of procedure in the educational network. Moreover, at the end, they are asked to fill in a questionnaire about the execution of the mechanism. The result of the questionnaire will be indicated in the result section\\
\hline
\end{tabular}
\end{center}
\label{tab:Implementation}
\end{table}

\section{Methodology}
\label{sec:methodology}
\begin{figure*}
	\includegraphics [width=\columnwidth]{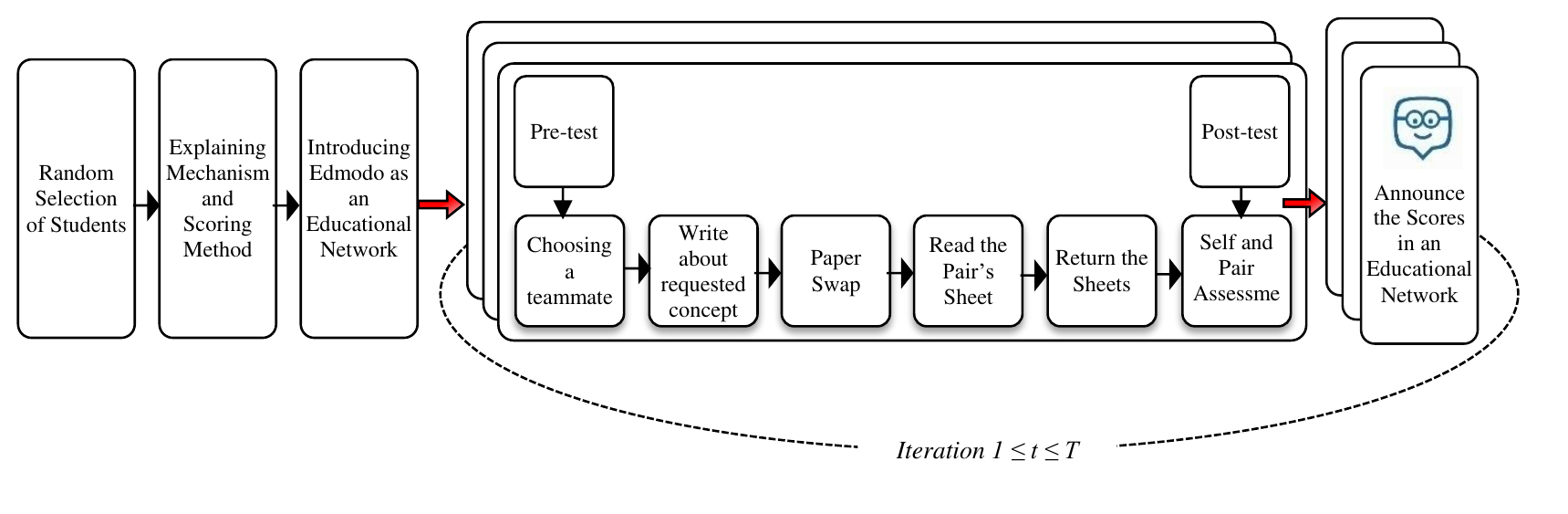}
	\centering
	\caption{Stages of PD\_PL implementation}
	\label{fig:PD_PL_Implementation}
\end{figure*}

In this section, we describe how to implement the mechanism and how to analyze the results. Figure \ref{fig:PD_PL_Implementation} shows the complete process of PD\_PL implementation.

At first, we asked the instructor to select a number of students randomly. Next, we explained the stages of PD\_PL and the scoring method to the selected students. We only told to them that our mechanism and scoring was set according to game theory. The majority of students had no knowledge about game theory, but some students knew about game theory through the movie \enquote{Beautiful Mind}.

In the first session, we taught how to register and enter the educational network in order to view the scores after each PD\_PL execution. In almost in every session that we ran PD\_PL, we described the scoring method. The user manual and the scoring method were also placed in the educational network. After each PD\_PL execution, top three scores of students in that session were also placed in the education network.

\begin{figure*}[h]
	\includegraphics [scale=0.95]{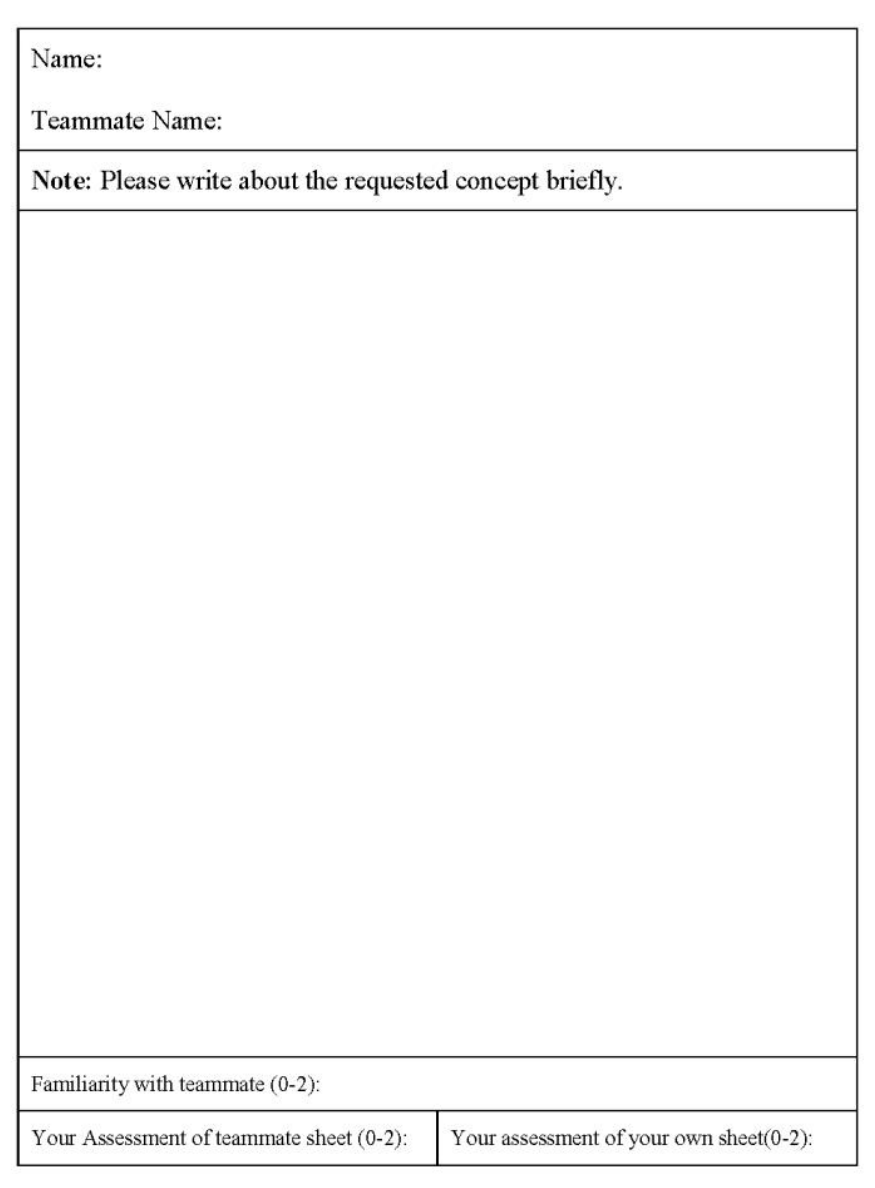}
	\centering
	\caption{A sample of sheet that students write about the requested concept}
	\label{fig:SampleSheet}
\end{figure*}

In each selected session, after the instructor taught all the students - including those chosen and those who were not chosen to participate in mechanism - we asked the selected students to stay in the classroom. Then, we asked the students to choose a teammate. The students were allowed to select a different teammate in different sessions. Then we handed the sheets to students and asked them to write about a requested concept briefly. Figure \ref{fig:SampleSheet} shows a sample sheet that we gave to students. The parts defined for assessment of the teammate's sheet, self-assessment, and familiarity with the teammate are observable at the bottom of the sheet.

\subsection{Participants}
 The participants were 142 students of Computer Engineering who enrolled in the courses \enquote{Fundamentals of Programming} (two different groups), \enquote{Discrete Mathematics}, and \enquote{Compiler principles techniques and tools}. We ran the proposed mechanism in the autumn and spring semesters of 2016. 

At the analysis stage, we would eliminate 23 students who participated in less than 3 sessions of PD\_PL execution. The information of remaining students is shown in Table \ref{tab:numberOfLearners}.

\begin{table*}[h]
\caption{The number of students who participate in more than two sessions of the mechanism as separated by course.}
\begin{center}
\begin{tabular}{c}
\includegraphics[scale=0.9]{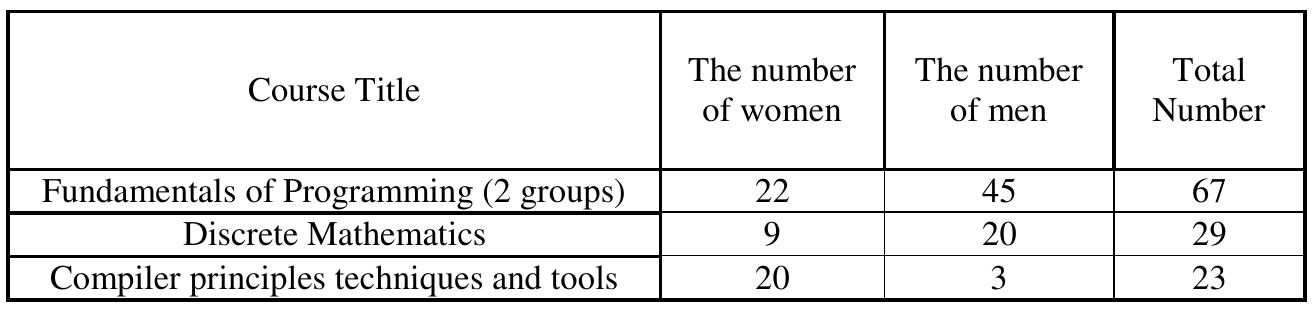}
\end{tabular}
\end{center}
\label{tab:numberOfLearners}
\end{table*}

For each selected course, we were going to run the mechanism in almost half of the sessions of a semester, so that the mechanism was run in seven different sessions of the semester.  As we chose four groups (Including two groups of \enquote{Fundamentals of Programming} and one group of each other courses), the mechanism was executed in 28 sessions. Table \ref{tab:selectedConcepts} shows the concepts used during these sessions in relation to each selected course.

\begin{table*}[h]
\caption{The concepts used in sessions in which the mechanism were executed.}
\begin{center}
\begin{tabular}{c}
\includegraphics[scale=0.8]{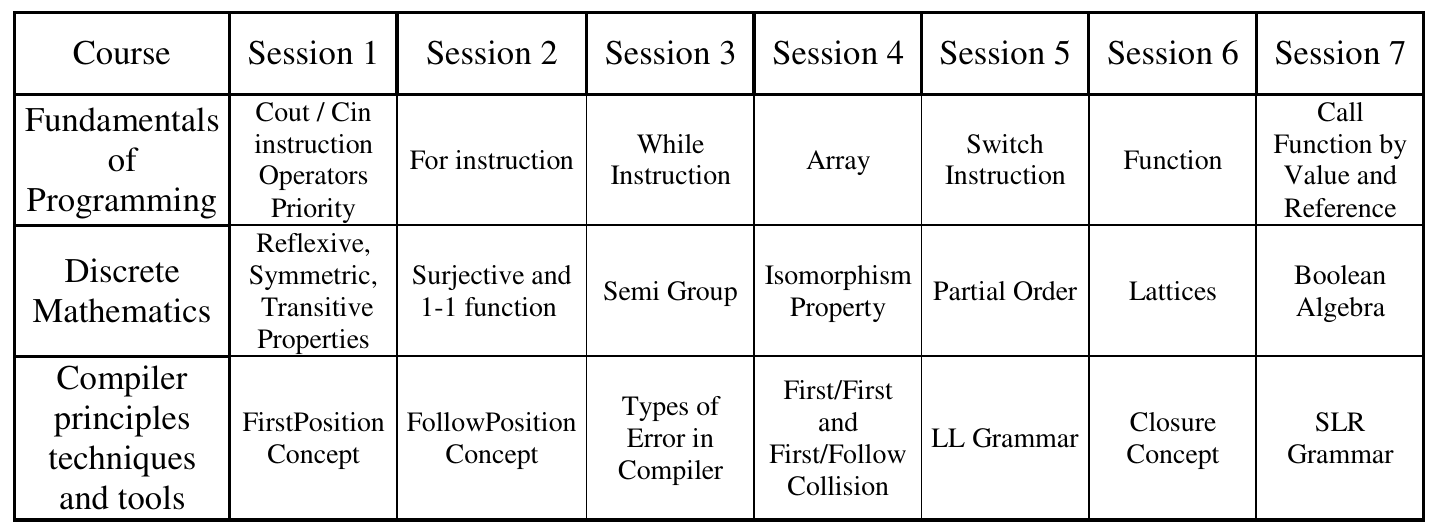}
\end{tabular}
\end{center}
\label{tab:selectedConcepts}
\end{table*}
 
\subsection{Data Analysis Method}

\begin{figure}
	\includegraphics [scale=1]{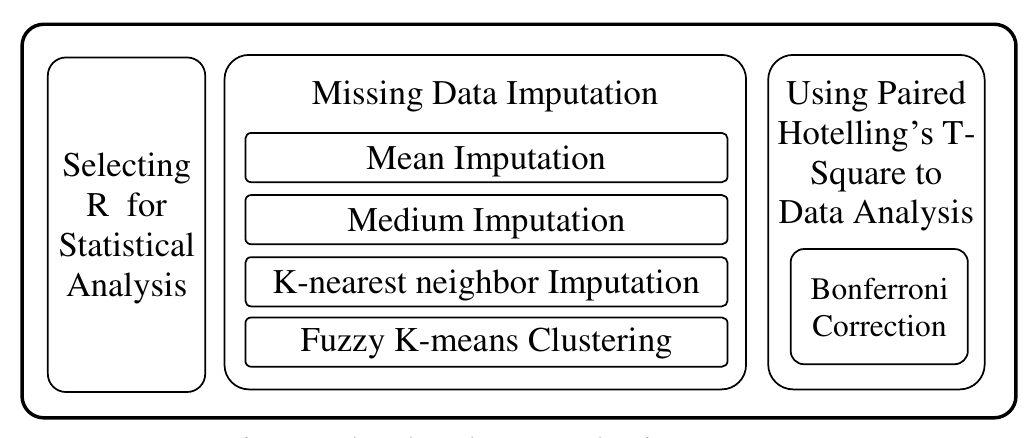}
	\centering
	\caption{The data analysis stages}
	\label{fig:AnalyzeMethod}
\end{figure}

Figure \ref{fig:AnalyzeMethod} shows the complete process of analysing our collected data. At first, the appropriate software to analyze the data is chosen. Then, as we used \enquote{Paired Hotelling's T-Square} to compare pre-test and post-test data and this test is unable to work with missing values, we explain the missing value imputation methods that we have used. Finally, using Paired Hotelling's T-Square, we investigated whether the proposed mechanism has a positive effect on learning.

A more detailed explanation of our analysis of the data collected is included below.

\subsubsection{Selecting R as a tool of data analysis}
To analyze the data, we use R programming language, which is known as the most widely used software in data mining and data science \cite{RN37}. R is an open source software. We used version 3.3.3 of this software as well as version 1.0.136 of RStudio, which is a graphical interface of R. 
\subsubsection{Missing Data Imputation}
As noted, the mechanism was executed throughout 7 sessions of each course. Some students were absent for some sessions. Therefore we faced some missing values, and to use Paired Hotelling's T-Square, missing data had to be imputed. So we apply four missing data imputation methods including using mean, using median, K Nearest Neighbour (KNN) and Fuzzy K-Mean Clustering (FKM). We used $R$ to program each method. For KNN, we used the package offered in article \cite{RN33}. 
   
FKM missing imputation is an efficient method \cite{RN22} that clusters data using an FKM algorithm and then fills in the missing data based on degrees of membership. In the FKM, each piece of data belongs to all the clusters with different degrees of membership. Algorithm \ref{alg:FKM} shows the FKM method in which each data $X_i (1 \leq i \leq N)$  includes \enquote{S} fields. In our work, $N$ equals to 119, which is the number of students.
 
As we execute our mechanism throughout the 7 sessions of each course, we have 7 pre-tests and 7 post-tests for each group. Therefore, each student has 14 scores in their post-tests and pre-tests, and so in our mechanism, the value of \enquote{S} is 14.

\begin{algorithm}
	\caption{Fuzzy K-Mean Clustering \cite{RN23}}
	\label{alg:FKM}
	\begin{algorithmic}[1]	
		\State Input $N$ data $X_i$, such that $(1 \leq i \leq N)$
		\State Set randomly $k$ initial centers as $V^{(0)}$.
		\State Set iteration number $r=0$
		\State $do$
		\State $\{$
	        \For {\textbf{each} $(1 \leq i \leq N)$} 
	        	\For {\textbf{each} $(1 \leq j \leq N)$}
	        		\State Calculate distance of $X_i$ from center of cluster $j$:
	        		\State $d(V_j,X_i)=(\Sigma_{i=1}^{n} |X_{is}-V_{js}|^p)^\frac{1}{p}$
	        		\State Calculate degree of membership of $X_i$ in cluster $j$:
	        		\State $U(V_j,X_i)=\frac{(d(V_j,X_i))\frac{-2}{m-1}}{\Sigma_{c=1}^{K}(d(V_c,X_i))\frac{-2}{m-1}}$
	        	\EndFor
			\EndFor
			\For {\textbf{each} $(1 \leq j \leq K)$}
			\State Update new centers of clusters:
			\State $V_{j}^{r+1}=\frac{\Sigma_{i=1}^{n} U(V_j,X_i) \times X_i}{\Sigma_{i=1}^{n} U(V_j,X_i)} $
			\EndFor
			\State $r=r+1$
			\State $\}while (\Vert V^{r+1}-V^r\Vert \geq\epsilon)$
		\end{algorithmic}
\end{algorithm}

The algorithm initially chooses $K$ data randomly out of the total data as K clusters' center (the second line of the algorithm). These centers are shown as $V^{(0)}$. Then, the distance of each piece of data from these $K$ centers is calculated (line 9). In the next step, the degree of membership of each piece of data in the centers of the K clusters is calculated (line 11). The value $m$ refers to a $Fuzzification Parameter$ that is usually equal to 2 \cite{RN23}. When a data includes some missing fields, we use $Partial Distance Strategy$ \cite{RN23} to calculate the distance:
\begin{equation}
\begin{aligned}
\label{PDS}
& d(X_i, V_k)= \dfrac{S}{I_i} \sqrt{2}{\Sigma_{j=1}^S(X_{ij}-V_{kj})^2I_{ij}}\\
& \mbox{Such that}\\
& I_{ij}=\left \{
\begin{array}{rcl}
0 & \mbox{if } X_{ij} \mbox{ is missing value}\\
1 & \mbox{otherwise}
\end{array}\right.\\
&I_{i}=\Sigma_{j=1}^S I_{ij} 
\end{aligned}
\end{equation}

Afterward, the new center of each cluster is defined using the weighted mean of the data existing in that cluster (line 16). The algorithm is repeated until the distance between two consecutive centers is higher than a defined threshold. When the distance is lower than the threshold, the algorithm will be stopped (line 19).

After clustering, the missing fields of data are imputed according to their degrees of membership. As an example, to impute the missing field $j$ of $X_i$, we can use the following equation.

\begin{equation}
\label{formula:Imputing}
X_{ij}=\frac{\Sigma_{k=1}^{K} U(V_k,X_i) \times V_{kj}}{\Sigma_{k=1}^K U(V_k,X_i)},
\end{equation}

Where, $V_kj$ refers to field j of the center of cluster k, and $U(V_k, X_i)$ refers to the degree of membership of $X_i$ in the center of cluster $k$. 

\subsubsection{Using Paired Hotelling's T-Square to Data Analysis}
Paired Hotelling's T-Square is an extended version of Paired t-test in multivariate situation. The Paired t-test is a statistical test that determines whether the mean difference between two groups of observations is zero. Suppose we are interested in evaluating the effectiveness of a training program. We can measure the performance of a sample of students before and after the training program, and examine the differences using a Paired t-test. 

The Paired t-test has two opposite hypotheses; the null hypothesis ($H_0$) and the alternative hypothesis ($H_1$). The $H_0$ assumes that the true mean difference between the paired samples is zero. The two-tailed $H_1$ assumes that the difference is not equal to zero.

Suppose $(x_1, y_1), (x_2, y_2),…. , (x_n, y_n)$ are pointing to a sample of a population where $x_i$ and $y_i$ refer to before and after observations. $H_0$ and $H_1$ are defined as below:

\begin{equation}
\begin{aligned}
H_0:\mu_x=\mu_y \\
H_1:\mu_x \neq \mu_y
\end{aligned}
\end{equation}

The values of $\mu_x$ and $\mu_y$ refer to mean of $x$ and $y$. If relation \ref{formula:t-square} is true, we reject the null hypothesis at signiﬁcant level $\alpha$ (with $\alpha$ chance of being mistaken) \cite{RN18,RN19}. The popular value of $\alpha$ are $0.1, 0.05$, or $0.01$. In our research, we set $\alpha$ to $0.05$.

\begin{equation}
T^2=\frac{(\bar{d})^2}{\frac{s^2}{n}}=n\bar{d}(s^2)^{-1}\bar{d} > t_{n-1}^2(\frac{\alpha}{2})
\label{formula:t-square}
\end{equation}

In relation \ref{formula:t-square}, $\bar{d}=\frac{1}{n}\Sigma_{i=1}^{n}d_i$  , where $d_i=x_i-y_i$ and $S^2=\frac{1}{n-1}\Sigma (d_i -\bar{d})^2$  is the variance of samples. The value of $t_{n-1}^2 (\frac{\alpha}{2})$ denotes the upper $100(\alpha/2)$th percentile of the t-distribution with $n-1$ degree of free.

As denoted, for each group, PD\_PL was executed in 7 sessions and on different concepts. If each concept is considered a variable, we face a multivariate problem. Therefore, we should investigate the effect of our mechanism on these different concepts by comparing the scores of pre-tests with post-tests.

In Paired Hotelling's T-Square, all scalar observations of the Paired t-test are replaced with vectors of observations. When post-tests and pre-tests are measured for the variables ($p$ variables, of which there are 7 in our mechanism), we compute vectors of differences ([post-tests]-[pre-tests]):

\begin{equation}
\begin{aligned}
&d_i=\left( \begin{array}{c} d_{i1} \\ \vdots \\ d_{ip} \end{array} \right)=\left( \begin{array}{c} x_{i1}-y_{i1} \\  \vdots \\ x_{ip}-y_{ip} \end{array} \right) \\
& for   1 \leq i \leq N
 \end{aligned}
\end{equation}

We calculate the value:

\begin{equation}
\label{tsquare}
T^2=n(\bar{d})' S^{-1} \bar{d}
\end{equation}

Such that:

\begin{equation}
\begin{aligned}
&\bar{d}_{(p\times 1)}=\frac{1}{n}\Sigma_{i=1}^n d_i \\
&S_{(p\times p)}=\frac{1}{n-1}\Sigma_{i=1}^n (d_i-\bar{d})(d_i-\bar{d})'
\end{aligned}
\end{equation}

If the relation \ref{eq:t-testcondition} is true, we reject the null hypothesis at signiﬁcant level $\alpha$.

\begin{equation}
\label{eq:t-testcondition}
T^2=n(\bar{d})' S^{-1} \bar{d} > \frac{(n-1)p}{(n-p)} F_{p,n-p}(\alpha)
\end{equation}

The value of $F_{p,n-p}(\alpha)$ denotes the upper 100($\alpha$/2)th percentile of the F-distribution with $p$ and $n-p$ degrees of free.

\subsubsection{Bonferroni Correction}
Any time we reject a null hypothesis, it is possible that we are wrong and the null hypothesis might be really be true, and our significant result might be coincidence. The $\alpha$ value of $0.05$ means that there is a $5\%$ chance of getting our observed result if the null hypothesis is true \cite{RN20}. 

The rejection of a true null hypothesis is called a type $I$ error. When we test multiple hypotheses in multivariate situations, $\alpha$ is the probability of making at least one Type $I$ error in multiple hypotheses. The Bonferroni correction is a classic method to solve the problem \cite{RN20,RN21} that tests each individual hypothesis at a significance level of $\alpha /m$ where $alpha$ is the chosen overall alpha level and $m$ is the number of hypotheses.

In our mechanism, we are going to compare seven pre-tests to seven post-tests. So we use $\alpha / 7$ instead of $\alpha$. 

\section{Results}
\label{sec:results}
In this section, we analyze the data of PD\_PL implementation and answer to research questions.

\subsection{Learning Improvement}
To answer research question 1, which was about our mechanism's effect on learning improvement, we use Paired Hotelling's T-Square. 

Table \ref{tab:prePost} shows a descriptive statistic of pre-tests and post-tests according to the 28 PD\_PL executions. The table shows the number of students, the mean, minimum score, maximum score, and the standard deviation of scores of each session, per course and in total. 

\begin{table}[hp]
\caption{The numbers of students, mean, minimum score, maximum score, and standard deviation of scores of each session per courses and in total.}
\begin{center}
\begin{tabular}{c}
\includegraphics[scale=0.8]{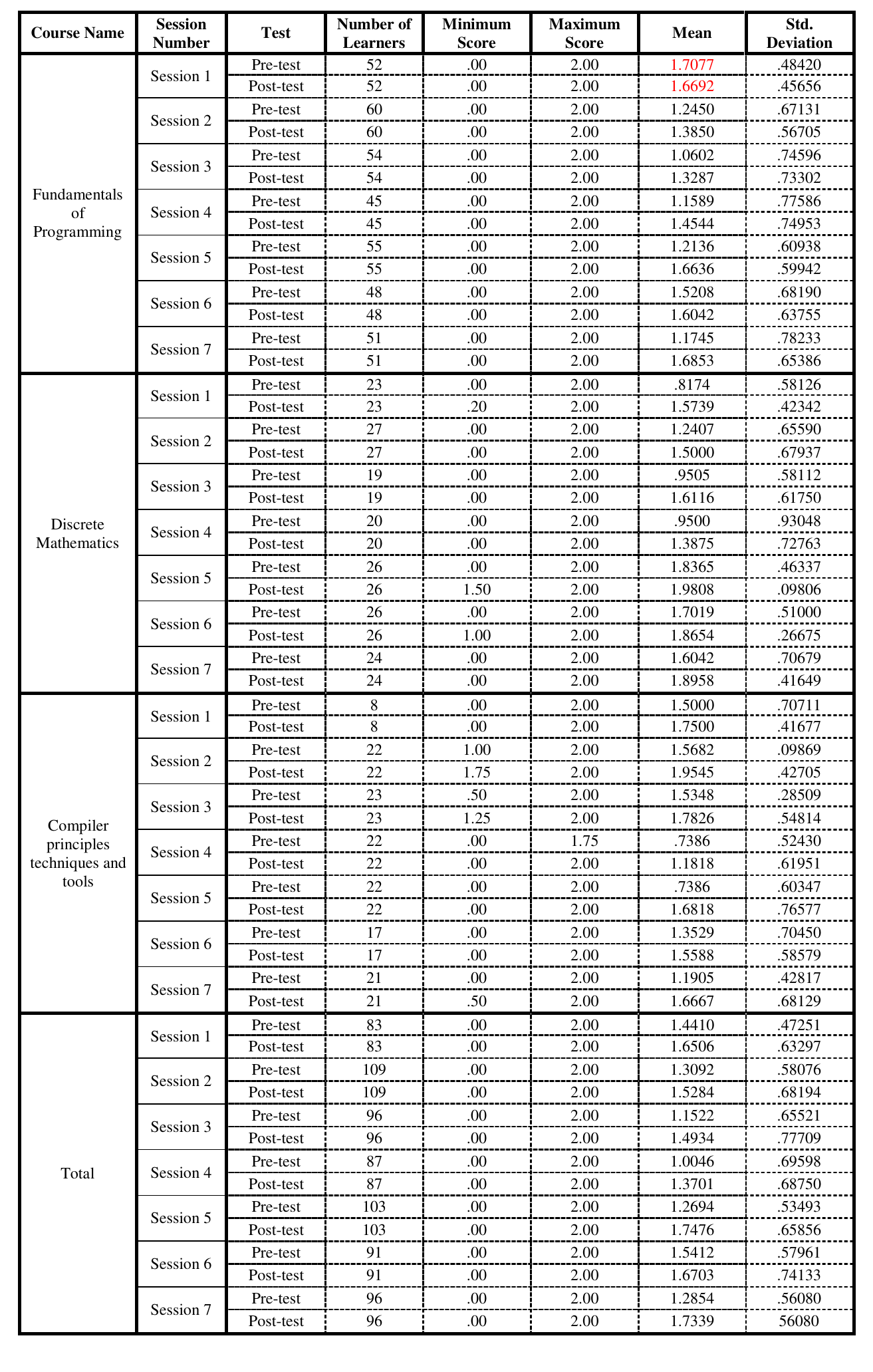}
\end{tabular}
\end{center}
\label{tab:prePost}
\end{table}

As Table \ref{tab:prePost} illustrates, the average scores of post-tests were higher than those of the pre-tests in all sessions; except for the first session of \enquote{Fundamentals of Programming}. This session was the first one in which our mechanism was run, and we asked the students to write about two relatively dense concepts. This exhausted the students, and consequently curtailed their performance. In other sessions, in consultation with the instructor, concepts that took between 5 to 10 minutes to be written about were chosen.
 
\begin{figure*}[hp]
	\includegraphics [scale=0.65]{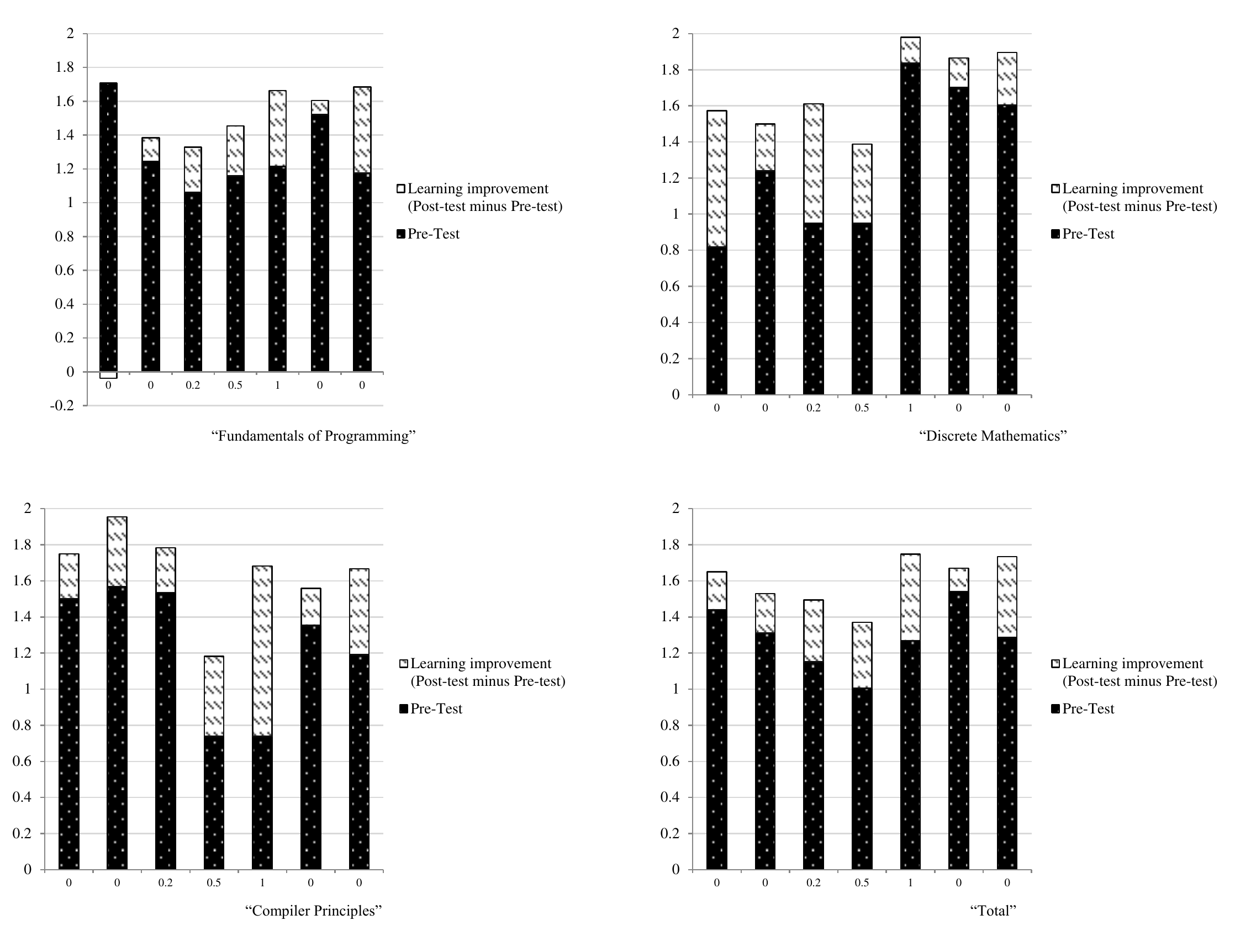}
	\centering
	\caption{The chart of difference between pre-tests and post-test per course and in total. The dark parts show pre-test and the light parts show difference between post-test and pre-test. In fact, the light part of each column shows the amount of learning improvement}
	\label{fig:prePostChart}
\end{figure*}

Figure \ref{fig:prePostChart} shows the difference between the post-test and the pre-test scores of each session per course and in total. 

%

As noted, using Paired Hotelling's T-Square we examined whether there were a significant difference between the pre-test and post-test scores. We used the missing data imputation as a requirement of Paired Hotelling's T-Square. With the $FKM$ method, as the first centers of clusters are randomly selected, the degree of membership of data (including data that contains missing fields) in the centers may be different. Therefore, we ran the $FKM$ missing data imputation algorithm three times.
 
At first, we calculated the value of $\frac{(n-1)p}{(n-p)}F_{p,n-p}\alpha$. The value of $n$ refers to the number of data that is equal to 119, and $p$ refers to number of comparison, which is equal to 7 in our test.
\begin{align*}
&\frac{(n-1)p}{(n-p)}F_{p,n-p}(\frac{\alpha}{7})=\frac{(119-1)7}{(119-7)}F_{7,11-7}(\frac{0.05}{7})\\
&=\frac{118 \times 7}{112} \times 2.948244=21.7433
\end{align*}

Table \ref{tab:tsquarevalue} shows the value of Hotelling's T-Square regarding Equation(\ref{tsquare}) after using different missing data imputation methods.

\begin{table}[t]
\begin{center}
\caption{The value of Paired Hotelling's T-Square after each of the mentioned missing data imputation methods}
\begin{tabular}{|c|c |} 
\hline
Method of missing data imputation   & $T^2$\\
\hline
Using Mean &	384.009\\
\hline
Using Median &	439.5052\\
\hline
KNN	& 365.9887\\
\hline
FKM (run 1) &	352.7527\\
\hline
FKM (run 2) &	353.3958\\
\hline
FKM (run 3) &	352.5303\\
\hline
\end{tabular}
\label{tab:tsquarevalue}
\end{center}
\end{table}

As shown in Table \ref{tab:tsquarevalue}, the value of Paired Hotelling's T-Square was higher than $\frac{(n-1)p}{(n-p)}F_{p,n-p}\alpha$ value in all situations. Therefore, we can reject $H_0$ and conclude that our mechanism (based on prisoner's dilemma and peer learning) enhances learning.

In another analysis of the pre-tests and post-test results, we found that the proposed mechanism could increase learning outcome from $4.17\%$ to $47.2\%$. We subtracted the mean value of post-test scores from the mean value of pre-test scores for each course and session. The column \enquote{Mean of post-test minus Mean of pre-test} in Table \ref{tab:prePostMeanDif} shows the results. The positive values show increasing scores in post-tests and consequent learning enhancement. Except for the first session of the \enquote{Fundamentals of Programming}, the pre-test scores were higher than the pre-test scores, which meant that PD\_PL had a positive impact on learning outcomes. The column \enquote{Percentage} of Table \ref{tab:prePostMeanDif} expresses the difference between the post-test and pre-test scores. The difference between the post-test and pre-test scores is expressed as a percentage. By ignoring the first session of \enquote{Fundamentals of Programming}, the minimum learning improvement is $4.17\%$ and the maximum is $47.2\%$.

\begin{table}[htp]
\caption{Difference between the mean of pre-test and post-test scores per course and session}
\begin{center}
\begin{tabular}{c}
\includegraphics[scale=0.7]{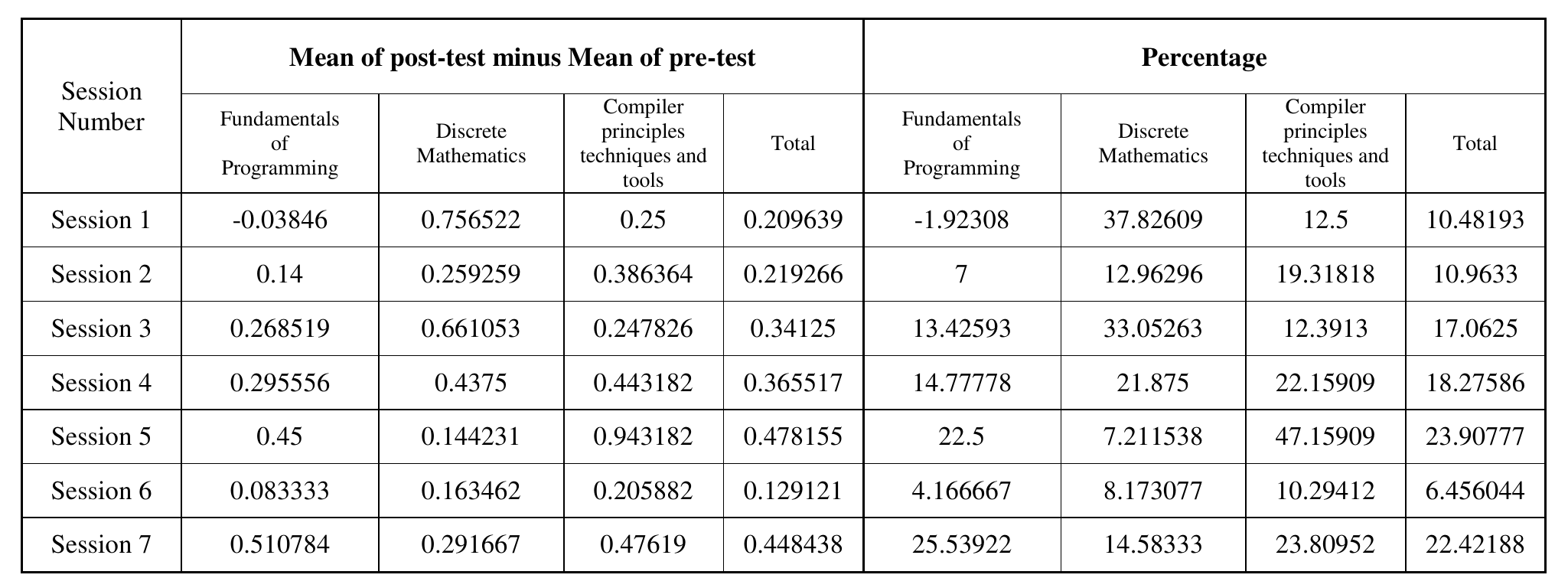}
\end{tabular}
\end{center}
\label{tab:prePostMeanDif}
\end{table}

\subsection{Free rider prevention}
The second research question was about our mechanism's ability to prevent free riding. Sometimes students may not know that they are doing less than the norm, therefore seeing the scores and what their peers are doing may encourage them to make more effort in future sessions \cite{freerider}. In PD\_PL, students have the opportunity to assess themselves and their peers on how much they do. In addition, at the end of each mechanism run, students could see teammates' and their own scores in the educational network.  
On the other hand, the students may really face free riders and have to do all the work. In this case, students may optimistically persuade their cohort to make more effort in subsequent sessions. They can even change their cohort in the next sessions to escape the free rider problem. 


Since students may choose different teammates in different sessions it is supposed that after several sessions, they can select a proper partner to collaborate in the learning process and subsequently obtain more payoffs. Given that some students may be absent from some sessions, we did some preliminary research to investigate how many of the groups were formed with the same members at the future sessions. We found in sessions 2 through 7 that $37.83\%, 41.05\%, 58.33\%, 59.78\%, 62.5\%$ and $64.9\%$ of the groups were formed by previous members for each respective session. For example, $37.83\%$ of the first session's teammates were also in a group in session 2.  An increase in the percentage groups' reformation shows that students gradually tended to choose one student as their cohort. Further research needs to be done to demonstrate effective parameters in group formation. We plan to develop this point in the next version of $PD\_PL$. 

%
\subsection{Subjective evaluation}
Students’ acceptance and usage are important measures of a mechanism's success. In early sessions of PD\_PL implementation, some students were worried about the amount of work that they had to do. Their initial impressions were that contributing in pre-tests and post-tests and filling in the sheets might be tedious. In later sessions, when they understood the positive effect of PD\_PL on their learning improvement, they had more motivation to participate in the mechanism. Interestingly, the teacher stated that during sessions in which the mechanism was not run, the students were following up and willing to run the mechanism.

\begin{figure*}[h]
	\includegraphics [scale=0.7]{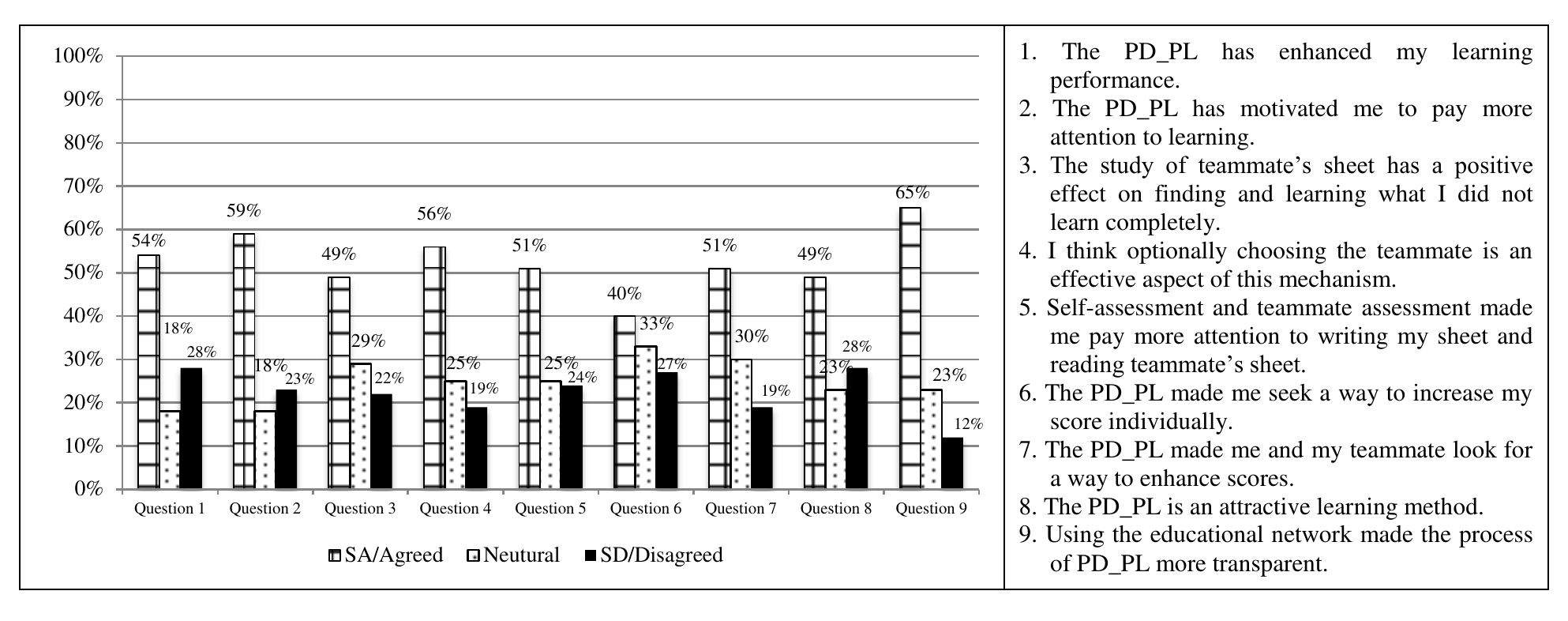}
	\centering
	\caption{Feedback of students on PD\_PL}
	\label{fig:feedback}
\end{figure*}

However in Section \ref{sec:ScoringMethod}, we stated some requirements of a peer learning environment. The last item mentioned was about giving the students feedback on PD\_PL. For this purpose, we prepared a questionnaire and asked students to fill it in. Figure \ref{fig:feedback} shows the questions and a representation of students' answers.

According to Figure \ref{fig:feedback}, some conclusions are drawn. For instance, the answers to questions 1 through 5 showed the positive effect of the mechanism on learning enhancement. Questions 6 and 7 investigated the mechanism from viewpoints of competition and cooperation, and responses suggested that the PD\_PL tended to increase rates of cooperation rather than competition. The answers to question 8 indicated that the mechanism was attractive for students. Finally, the outcome of question 9 indicated that a facility like the educational network had a positive influence on the mechanism's performance.

\section{Discussion}
Implementation of the mechanism faced several challenges. For example, as Table \ref{tab:prePost} shows, the average scores of post-tests were higher than those of the pre-tests in all sessions except the first session of \enquote{Fundamentals of Programming}. This session was the first one during which PD\_PL was run, and the instructor asked the students to write about two relatively dense concepts. This tired the students, and consequently reduced their performance. In other sessions, in consultation with the instructor, educational concepts that took between 5 to 10 minutes to be written about were selected. As another example, we can mention the preparation of the implementation requirements. Preparing the pre-test and post-test questions, correcting them, inserting the scores into the educational network, and monitoring its implementation and directing the students were a very difficult and time consuming tasks for the researcher and the instructor of the course.

\section{Conclusion and future works}
\label{sec:Conclusion}
In this study, similarities in they situations of peer learning and prisoner's dilemma was used to propose PD\_PL as a game theoretical approach to peer learning.

The proposed mechanism was implemented during several sessions with 142 students. The results of pre-test and post-test exams of for all the sessions were compared using R software through Paired Hotelling's T-Square analysis to investigate the impacts of $PD\_PL$ and the proposed instructional design on students' personal learning. As Paired Hotelling's T-Square is not able to function with missing data, we applied four different missing data imputation methods including using mean, using median, K Nearest Neighbour (KNN) and Fuzzy K-Mean Clustering (FKM). We also used Bonferroni correction to solve the type $I$ error in multivariate situations. 

The preliminary evaluation indicated that the mechanism had a positive effect on the the learning enhancement. It may be interpreted that PD\_PL could propose an acceptable mapping between a prisoner's dilemma atmosphere and peer learning since in PD\_PL, the efforts of teammates resulted in a higher payoff for both of them and consequently increased learning outcomes. 
Since PD\_PL lets students see their teammates efforts, they might escape the free rider problem by changing the teammate or persuading them to be more active.
In addition, the results of a subjective evaluation revealed that the majority of students found PD\_PL to be an attractive and efficient tool for learning enhancement. 
The most important findings are:
\begin{itemize}
\item Learning Improvement: Putting the students together and hopping for the best is not appropriate peer learning implementation. PD\_PL prepared a peer learning environment using prisoner's dilemma. It passed the preliminary verification process and had a positive effect on learning enhancement.
\item Free rider prevention: At its mostly poorly designed, peer learning may for instance result in one person making all the effort. The ability to see the teammate's sheet and final score in the educational network and being able to chang peers in different sessions is a way to prevent the free rider problem.
\item Subjective evaluation: The result of subjective research showed the mechanism to be attractive. It also indicated that PD\_PL tended towards cooperation in the learning process.
\end{itemize}

We encountered some limitations with PD\_PL implementation. As mentioned, the mechanism had been run during several sessions and the absence of some students for some sessions made missing data in the pre-tests and post-tests data. However, as described in the methodology section, we minimized the effect of the missing data by using a missing data imputations algorithm. To increase accuracy, we also used four imputation methods in parallel. Another problem we encountered was in cases the number of selected students in a session was odd. To solve the problem, we told the odd-one out student, who was alone and did not have a cohort, to participate alone. We ignored this student at the analysis stage.

The proposed mechanism is a general method that can be implemented in any blended environments that possess interaction and scoring ability. The instructional design can also use this mechanism - even in custom classrooms.
 
More work is yet to be done to determine the teammate changing pattern during different sessions. The research has raised many unanswered questions; for example, whether the teammates' behavior in previous sessions impacted the amount of knowledge written on the sheet in the next session. There are different methods to choosing strategy in the iterated prisoner's dilemma \cite{RN34}. As another interesting point, we aim to investigate students' behavior during sessions of PD\_PL implementation regarding the mentioned methods in reference \cite{RN34}.
\newpage
\bibliographystyle {apa}

\bibliography{mybib}{}
%

\end{document}